# Spectroellipsometric and ion beam analytical studies on a glazed ceramic object with metallic lustre decoration

T. Lohner*[1], E. Agócs[1], P. Petrik[1,2], Z. Zolnai[1], E. Szilágyi[3], I. Kovács[3], Z. Szőkefalvi-Nagy[3], L. Tóth[1], A.L. Tóth[1], L. Illés[1], I. Bársony[1,2]

[1]Research Centre for Natural Sciences, Institute for Technical Physics and Materials Science, Hungarian Academy of Sciences, H-1121 Budapest, Konkoly Thege Miklós út 29-33, Hungary

[2]Doctoral School of Molecular- and Nanotechnologies, Faculty of Information Technology, University of Pannonia, Egyetem utca 10, Veszprém, H-8200, Hungary

[3]Institute for Particle and Nuclear Physics, Wigner Research Centre for Physics, Hungarian Academy of Sciences, H-1121 Budapest, Konkoly Thege Miklós út 29-33, Hungary

*Corresponding author: lohner@mfa.kfki.hu

**Abstract**

In this work recently produced and commercially available glazed ceramic object with metallic lustre decoration was studied by using a spectroscopic ellipsometer with rotating compensator. The thickness and metal content of the surface lustre layers are determined by ion beam analytical techniques, i.e., Rutherford backscattering spectrometry and external beam particle-induced X-ray emission and the results were utilized in the construction of multilayer optical models for the evaluation and interpretation of the spectroellipsometric measurements.

## 1. Introduction

In the 9th century in Mesopotamia a special technique for the decoration of ceramic objects appeared - lustre, a precursory nanotechnology, which is able to transform simple earth into ceramic masterpieces, thus giving them beautiful metallic shine, including the appearance of gold [1]. Lustre decoration of medieval and renaissance pottery consists of silver and copper nanocrystals, dispersed within the glassy matrix of the ceramic glaze [2]. Roqué et al. aimed to establish a basis for understanding lustre nanostructure linked to its optical properties [3]. Pradell et al. performed Rutherford backscattering spectroscopy (RBS) and optical measurements on laboratory prepared lusters and they have found that the high volume fraction of metal nanoparticles is responsible for the metallic shine [4]. Recently, a metallized glaze has been produced from sepiolite-n(Cu, Fe) containing metallic nanoparticles and applying a fast-firing fabrication process by Moya et al. [5]. Based on ellipsometric and reflectance measurements, Moya et al. draw the conclusion that copper nanoparticles in the topmost glaze layer is a prerequisite to obtain a metallized glaze.

Particle-induced X-ray emission (PIXE) allows for the fast and simultaneous identification of a large number of elements with reasonable accuracy [6] whereas RBS is a widely used method for the surface layer analysis of solids [7] with good depth resolution. Polvorinos del Rio et al. studied lustre glazed ceramics from the medieval Seville by PIXE and RBS [8]. Their simulation of RBS spectra shows the existence of thin layers containing metallic silver and/or copper. One of the early examples of lustred object produced in Italy is Baglioni's albarello, this emblematic object was studied by Padeletti et al. using PIXE and RBS among other methods to gain deeper insight into the characteristics of the lustred film [9].

The processing route to obtain a glazed ceramic object with lustre decoration consists of three steps: i) high temperature firing of the ceramic body (< 1000 $^{o}$C), ii) application and firing of the vitreous glaze at intermediate temperature (500 - 900 $^{o}$C), iii) application of a special mixture containing clay, metallic (Ag, Cu) salts and organic compounds (for example vinegar) and firing in reducing atmosphere at moderate temperature (600 $^{o}$C). During this latter treatment metallic nanoparticles are formed which remain embedded in a thin near-surface layer of the ceramic object [1].

In this work a recently produced commercially available heart-shaped glazed ceramic pendant with metallic lustre decoration was studied by spectroscopic ellipsometry (SE). The thickness and metal content of the surface lustre layers were determined by ion beam analysis, i.e., RBS and PIXE. The results of ion beam analysis helped us to construct multilayer optical models for the evaluation and interpretation of the SE measurements. The photograph in Fig. 1 shows the heart-shaped ceramic pendant together with two coins for comparison in size.

## 2. Experimental details

2.1 Ion beam analysis

The surface of the recently produced, commercially purchased heart-shaped ceramic pendant with metallic lustre decoration has a certain degree of curvature. Considering this, to minimize the uncertainty of the applied measurement geometry (i.e. actual sample tilt angle, layer thicknesses, etc.) in the spectrum evaluation process, the areas for ion beam analysis and optical investigations were selected on surface regions of minimum curvature.

The external beam PIXE and RBS analysis were performed using the 5 MV Van de Graaff accelerator at the Institute for Particle and Nuclear Physics, Wigner Research Centre for Physics, Hungarian Academy of Sciences, Budapest.

In case of external beam PIXE, the pin-hole collimated proton beam of 2.5 MeV energy was extracted to air through a 7.5 μm thick Kapton foil. The sample holder with the ceramics used in the RBS measurements were simply set to face to the extracted beam. The final target positioning was achieved using a mechanical "aiming pin pointer", see the photograph of the measurement setup in Figure 2. X-ray spectra were collected by a computer controlled Amptek X-123SDD spectrometer of 25 mm$^2$ x 0.5 mm active detector volume, 8 μm thick Be window and 130 eV energy resolution for the Mn K$_\alpha$ line. In front of the detector an Al absorber of 0.1 mm thick was used. The net X-ray peak intensities were evaluated with the GUPIX program package [10]. The overall sensitivity of the setup is in the 10-50 ppm range.

For RBS, the sample Z1 was fixed to a sample holder of a scattering chamber equipped with a two-axes goniometer. During the experiments the vacuum in the chamber was better than $1 \times 10^{-4}$ Pa using liquid N$_2$ traps along the beam path and around the sample. The ion beam of 2000 keV $^4$He$^+$ was collimated with 2 sets of four-sector slits to the necessary dimensions of 0.5 x 0.5 mm$^2$. The ion current of typically 10 nA was kept constant via monitoring by a transmission Faraday cup [11]. The spectra were collected with a measurement dose of 4 μC.

The RBS measurements were performed with an ORTEC surface barrier detector under a solid angle of 4.15 msr. The energy calibration of the multichannel analyzer (2.99 keV/channel resolution and with energy offset of 104.7 keV) was performed by using known peaks and surface edges of Au, Si and C, respectively. To identify the surface elements and buried peaks, the RBS experiments were performed at tilt angles of 7$^o$ and 45$^o$, respectively. To determine the lateral homogeneity of the sample RBS experiments were repeated on three different spots of the heart-shaped ceramic pendant. The three spectra taken on the sample were then fitted using the layer structure obtained from the RBX simulation program [12].

2.2. Spectroscopic ellipsometry

The optical properties and the thicknesses of thin film structures can be derived from ($\Psi,\Delta$) values measured by SE, where $\Psi$ and $\Delta$ describe the relative amplitude and relative phase change of polarized light during reflection, respectively. In the present experiment $\Psi$ and $\Delta$ were measured by a Woollam M-2000DI rotating compensator ellipsometer in the 191 – 1690 nm wavelength range at angle of incidence of $70^o$. A 0.3 mm diameter microspot option was applied in the measurement. The calculated (generated) spectra were fitted to the measured ones using a regression algorithm. The measure of the fit quality is the mean square error (MSE) which was compared for different optical models. The unknown parameters are allowed to vary until the minimum of MSE is reached. Since the regression algorithm may end up in a false "local" minimum, therefore a careful global search procedure should be applied in case of complex multilayer structures. That means that one should start the evaluation with a wide range of initial parameter values in order to find the global minimum. For complex sample structures this procedure can be time consuming.

**3. Results and discussion**

Figure 3 shows a typical PIXE spectrum measured on the heart-shaped ceramic pendant. The X-ray lines corresponding to various elements with $Z \geq 26$ (i.e., Fe and above) are identified. The relative amount of the various metal components can easily be calculated from the peak areas. Significant amounts of Fe, Cu, Zn, Zr, Ag, Sn and Pb were found in the surface layer.

The depth distribution of the above elements was determined by RBS. Figure 4 shows the RBS spectra taken on the center part of the heart-shaped ceramic pendant along with the simulated ones. Besides the metallic elements found by PIXE O, Si and Na were also detected in the evaluation. To obtain a good quality agreement between measured and simulated RBS spectra about fifteen different layers of various composition and layer thickness were introduced. Basically, the ceramic matrix, which can be considered as the substrate, was formed by the elements of Si, O, Na, Fe, Cu, Zn, Sn and Pb. Although the energy spread effects were calculated in the simulation [13], several layers had to be introduced to describe the shape of the Zr and Ag peak of various concentration. The concentration of Sn and Pb was also found to vary with depth.

Comparison of the spectra taken on various spots on the sample suggests the presence of some inhomogeneity in the lustre layer. The thicknesses of the Zr and Ag containing layers, as well as that of the third matrix layer slightly vary.

We have to note that PIXE and RBS result in different element ratios for Zr and Ag. While PIXE gives for the ratio of Zr:Ag = 0.1:1, RBS results in 0.9:1. This disagreement can be solved if we take into account that the analysed depth is quite different for both methods (10-20 µm for PIXE, and ca. 2 µm for RBS). The discrepancy can be resolved assuming another Ag containing layer below a depth of 2 µm. Another possibility is that the ceramic matrix itself contains some Ag constituent.

The layer thicknesses determined by RBS are given in units of at/cm$^2$. The layer thicknesses were converted to nm scale using an atomic density of $6.67 \times 10^{22}$ at/cm$^2$ in order to facilitate the development of the optical model. Table 1 displays the results of the simulations for the lustre RBS spectra. Here the concentrations are given in at%. In Table 1, the most remarkable feature is the presence of buried Ag and Zr with maximum concentrations at depths of about 120 nm and 190 nm, respectively, in addition to the multicomponent substrate.

Only one paper described spectroellipsometric study on nanostructured metallized glazes so far [5]. Those authors applied a double-layer optical model with Cu, CuO and SiO$_2$ components.

In our case a multilayer optical model with Lorentz oscillators and reference data from literature (Ag [14], SiO$_2$ and ZrO$_2$ [15]) was applied for the evaluation of the ellipsometric spectra. Evaluation was carried out using the WVASE32 software of the J.A. Woollam Co., Inc.

Two Lorentz oscillators were introduced to model the materials of which the glaze is composed. 15 unknown parameters (oscillator parameters, layer thicknesses, volume fractions of metallic silver, silicon dioxide and zirconium-oxide) were defined and fitted.

Based on the results from PIXE and RBS, about 50 different optical models of increasing complexity were conceived and used for evaluation of the measured ellipsometric spectra. Although fifteen different layers of various composition and thickness were introduced to obtain good quality match between measured and simulated RBS spectra, in case of the SE optical model the number of layers and unknown parameters was limited to 4 and 15, respectively, to avoid cross-correlation. Figure 5 shows the multilayer optical model for the evaluation of spectroellipsometric data. Since there is no hope to find reference dielectric functions for the materials of the sublayers of the ceramic matrix consisting of about ten different elements, we decided to apply general oscillators for optical modelling. The dielectric function of the substrate has been described by a Lorentz oscillator (Lorentz-1). The optical properties of the layer next to the substrate (layer-1) has been represented by another Lorentz oscillator (Lorentz-2). Each Lorentz oscillator involves four free parameters to be determined during the evaluation of the measured spectra. The layer adjacent to layer-1 (layer-2) was supposed to be a mixed layer with constituents of the material of layer-1 and $ZrO_2$ by taking into account the RBS results. The dielectric function of the mixed layers was calculated on basis of effective medium approximation [16, 17]. Similarly, layer-3 was supposed to be a mixed layer with constituents of the material of layer-1 and Ag by taking into account the RBS results. Layer-4 was represented as a mixture of the material of substrate and $SiO_2$.

Figure 6 and 7 shows the measured and generated ellipsometric $\Psi$ and $\Delta$ spectra for the glazed ceramic piece with greenish-gold metallic lustre decoration. The multiparameter evaluation yielded 97.2 for the mean square error. There is a satisfactory agreement between the measured and generated spectra.

Figure 8 displays the multilayer optical model with the results of evaluation of SE data: the volume fractions of the components and the layer thicknesses together with their confidence intervals are shown. The reason for the obtained high volume fraction of $ZrO_2$ (99.1%) might be the high refractive index of the material in the vicinity of the relatively high concentration of Zr.

Evaluation based on a five-layer optical model containing a supplementary middle layer (between the actual layers 2 and 3) containing both components Ag and $ZrO_2$ as resulted from RBS simulation (Table 1, layer number 9) resulted in MSE=94.04. This value is somewhat lower than the MSE value (97.2) of the evaluation based on the previous (four-layer) optical model. According to the evaluation based on the five-layer optical model the composition of the supplementary middle layer is as follows: the material described by a Lorentz oscillator (Lorentz-2) with volume fraction of 27%, $ZrO_2$ with volume fraction of 62 ± 19% and Ag with volume fraction of 0.8 ± 1.3%. The low percentage and the uncertainty (0.8 ± 1.3%) of Ag volume fraction does not indicate the presence of Ag in the supplementary middle layer. The evaluation gave a thickness value of 6 ± 2 nm for the supplementary middle layer.

The direct comparison of the thicknesses and positions of Ag-rich and Zr-rich layers deduced from RBS and SE results may be not straightforward mainly because of the uniform atomic density used to calculate all the layer thicknesses, however, if we consider the depth distribution of elements yielded by RBS (Table 1) and the layer structure and composition based on results from ellipsometric evaluation (Figure 8) a certain kind of similarity can be found. One fact is that both methods show that the center of the silver depth distribution is closer to the sample surface than that of the zirconium depth distribution.

Preliminary cross-sectional electron microscopy investigations show the presence of spherical nanoparticles with diameter in the range from 5 to 50 nm, high resolution electron microscopy images proved that there are silver nanoparticles in the specimen showing 111 and 200 lattice planes of silver.

**Conclusions**

Spectroscopic ellipsometry, particle induced X-ray emission and Rutherford backscattering spectrometry were employed to investigate the surface lustre layers of a commercially purchased ceramic pendant. A four-layer optical model based on effective medium approximation and on Lorentz oscillators proved to be the appropriate evaluation method yielding satisfactory agreement between the measured and calculated ellipsometric data. From the experiments we have extracted the layer thicknesses and information concerning composition of these layers. Our analysis confirmed the assumption that Ag nanoparticles might be responsible for the metallic shiny greenish-gold appearance of the ceramic pendant [4].

**Acknowledgements**

RBS and PIXE experiments were carried out in the frames of the Hungarian Ion-beam Physics Platform. The support of the Hungarian grants TÁMOP-4.2.2/B-10/1-2010-0025 and OTKA K81842 is acknowledged.

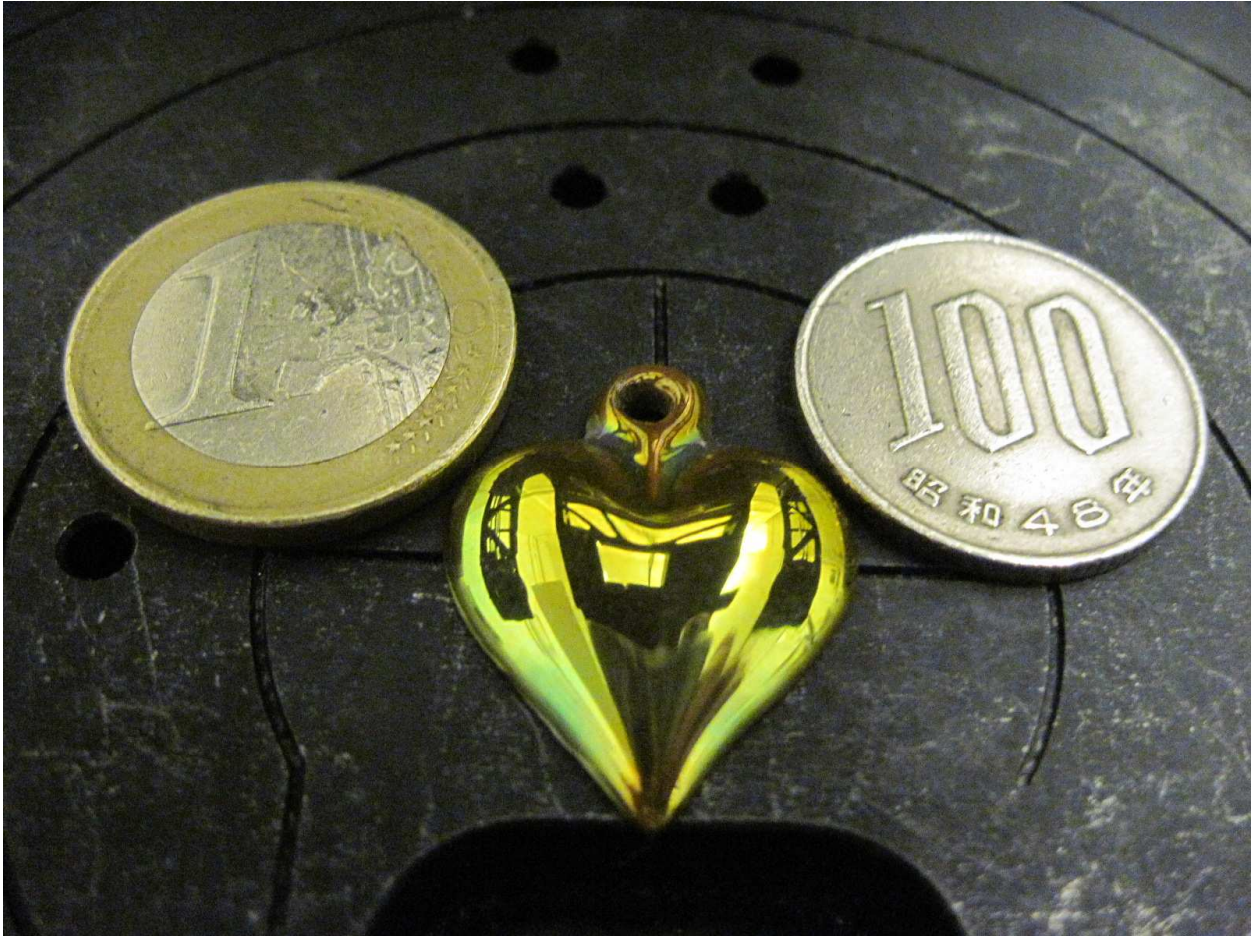

Figure 1. Photograph of the commercially purchased heart-shaped ceramic pendant with metallic lustre decoration.

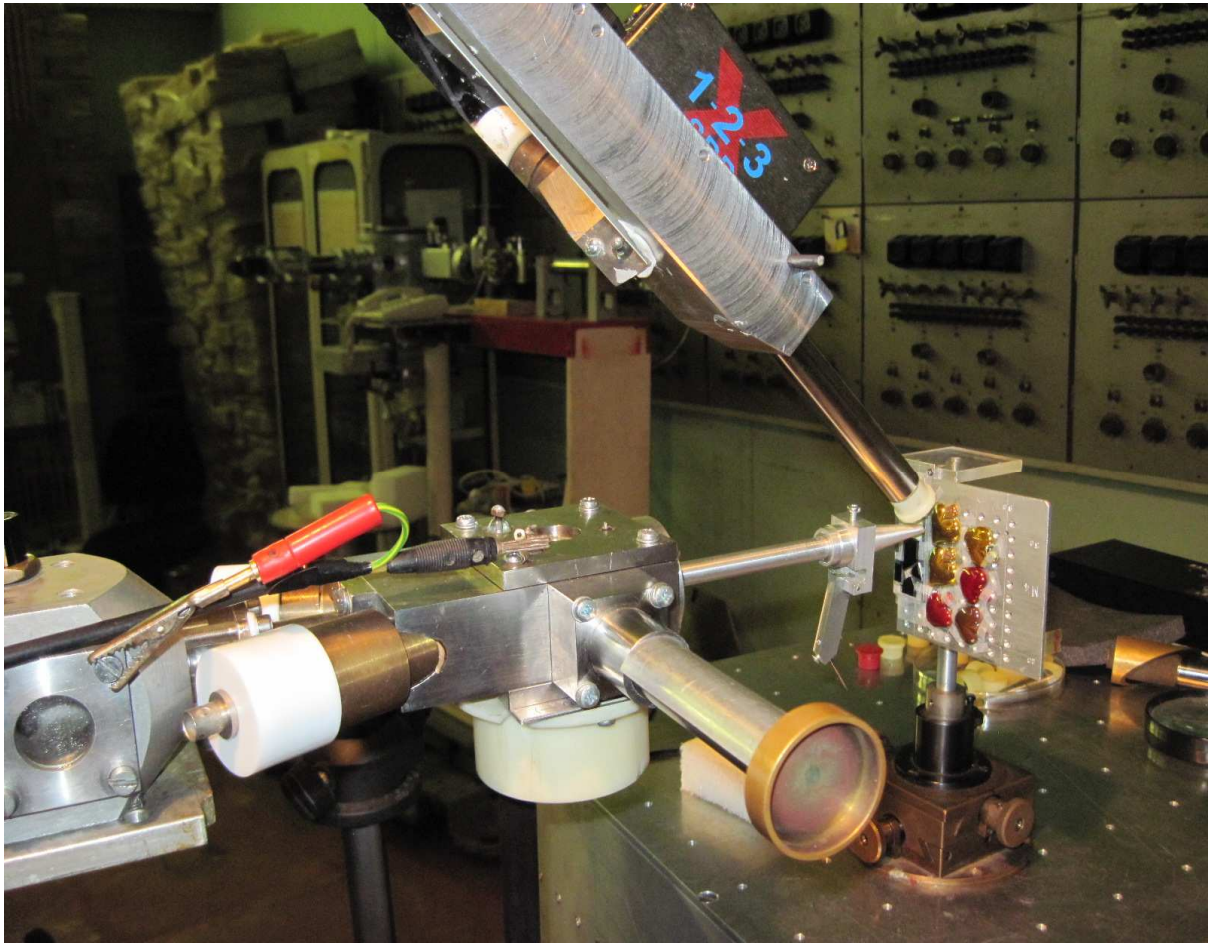

Figure 2. The external beam PIXE setup.

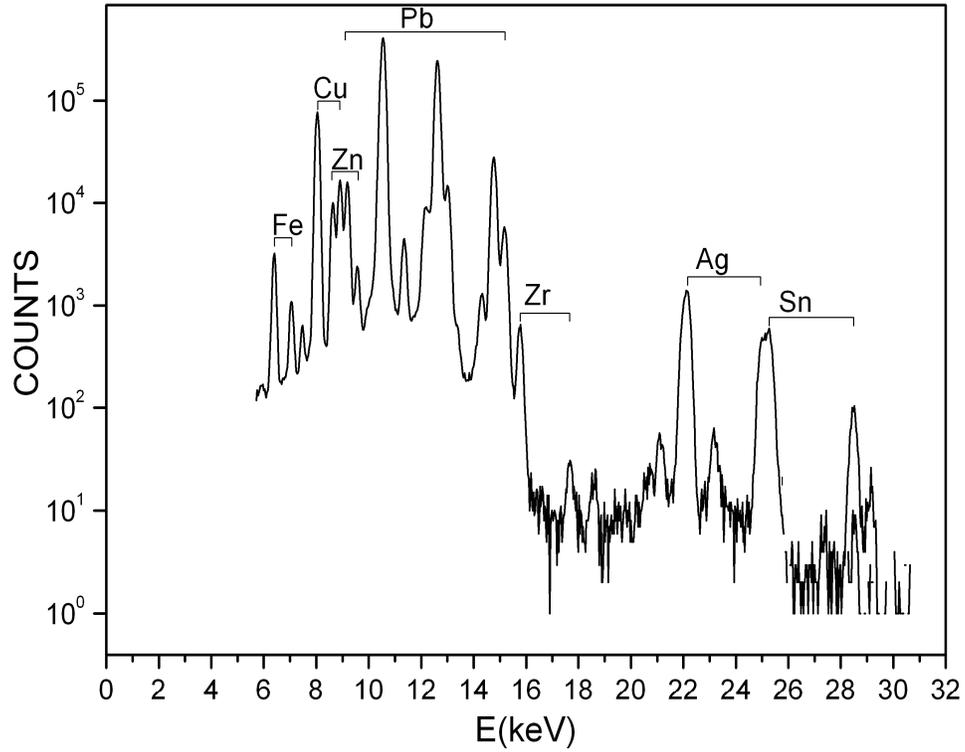

Figure 3. Typical PIXE spectrum measured on the heart-shaped ceramic piece with metallic lustre decoration. The X-lines of elements with atomic number Z ≥ 26 were analyzed.

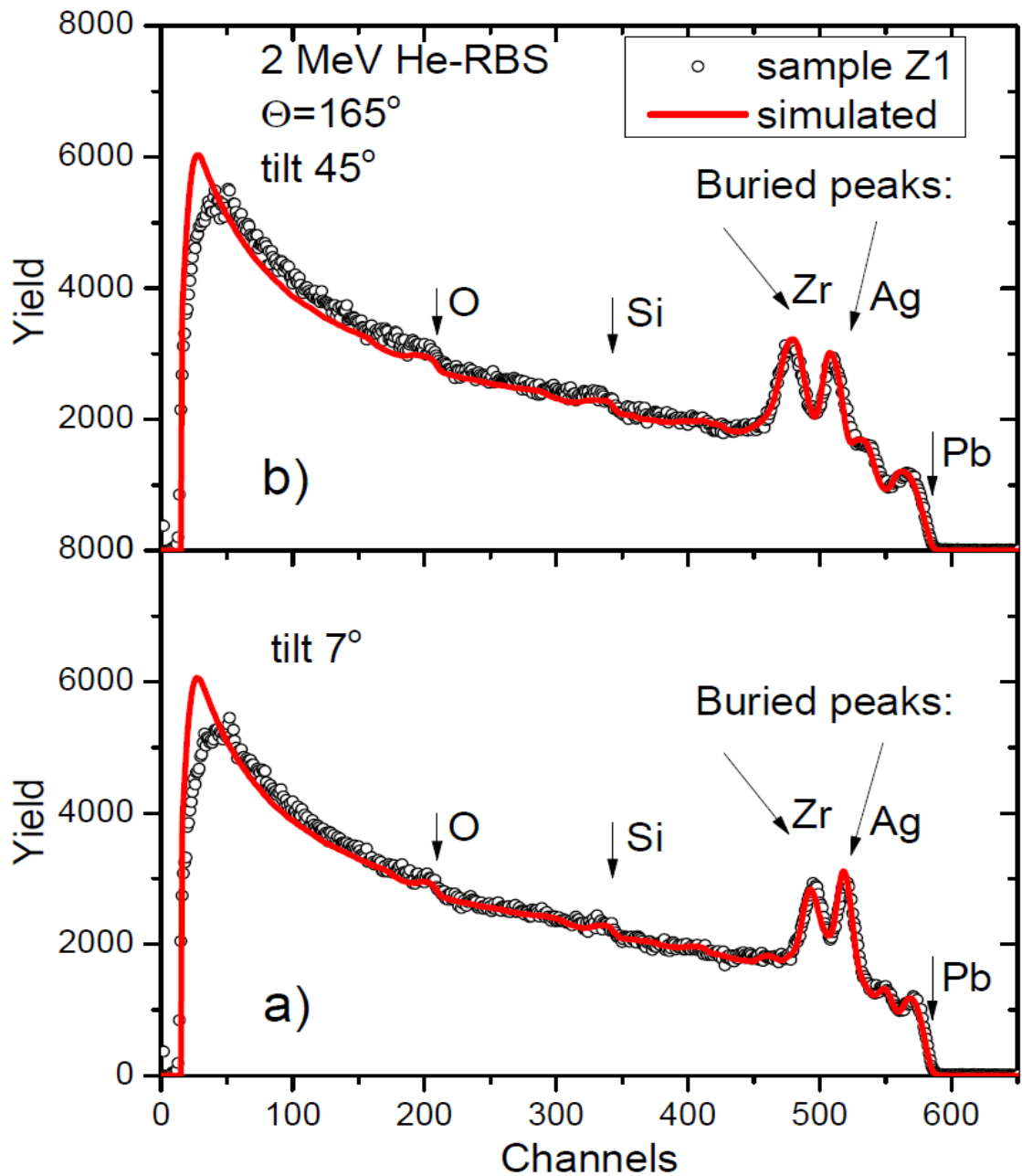

Figure 4. Measured and simulated RBS spectra taken on the heart-shaped ceramic pendant with metallic lustre decoration at a) tilt 7° and b) 45°. The surface position of O, Si, and Pb edges was denoted by arrows. The two peaks corresponding to buried Zr and Ag are also indicated.

Table 1. Results of the simulations for the lustre RBS spectra, the concentrations are given in at%. (An atomic density of 6.69×10$^{22}$ at./cm$^3$ used to calculate the layer thicknesses in units of nm.)

| layer numbers | Si | O | Na | Ca | Fe | Cu | Zn | Sn | Pb | Ag | Zr | layer thickness (at./cm$^2$) | thickness (nm) |
|---|---|---|---|---|---|---|---|---|---|---|---|---|---|
| 15 | 20 | 61 | 4 | 5 | 4 | 2 | 2 | 0 | 1 | 0 | 0 | 6.00×10$^{16}$ | 9.0 |
| 14 | 20 | 59 | 4 | 5 | 4 | 2 | 2 | 0 | 2 | 3 | 0 | 1.60×10$^{17}$ | 23.9 |
| 13 | 19 | 58 | 4 | 5 | 4 | 2 | 2 | 0 | 3 | 3 | 0 | 1.34×10$^{17}$ | 20.0 |
| 12 | 19 | 56 | 4 | 5 | 4 | 2 | 2 | 0 | 4 | 5 | 0 | 2.47×10$^{17}$ | 36.9 |
| 11 | 15 | 45 | 3 | 4 | 3 | 1 | 2 | 0 | 5 | 19 | 3 | 6.69×10$^{16}$ | 10.0 |
| 10 | 15 | 46 | 3 | 4 | 3 | 2 | 2 | 0 | 3 | 20 | 3 | 1.34×10$^{17}$ | 20.0 |
| 9 | 13 | 38 | 3 | 3 | 3 | 1 | 2 | 0 | 4 | 25 | 9 | 1.36×10$^{17}$ | 20.4 |
| 8 | 15 | 44 | 3 | 4 | 3 | 1 | 2 | 0 | 5 | 4 | 19 | 3.00×10$^{17}$ | 44.8 |
| 7 | 19 | 57 | 4 | 5 | 4 | 2 | 2 | 0 | 3 | 2 | 2 | 4.50×10$^{17}$ | 67.3 |
| 6 | 19 | 57 | 4 | 5 | 4 | 2 | 2 | 2 | 4 | 0 | 1 | 4.00×10$^{17}$ | 59.8 |
| 5 | 19 | 57 | 4 | 5 | 4 | 2 | 2 | 2 | 5 | 0 | 0 | 3.00×10$^{17}$ | 44.8 |
| 4 | 19 | 58 | 4 | 5 | 4 | 2 | 2 | 2 | 4 | 0 | 0 | 3.00×10$^{17}$ | 44.8 |
| 3 | 19 | 57 | 4 | 5 | 4 | 2 | 2 | 3 | 4 | 0 | 0 | 3.00×10$^{17}$ | 44.8 |
| 2 | 19 | 58 | 4 | 5 | 4 | 2 | 2 | 2 | 4 | 0 | 0 | 3.00×10$^{17}$ | 44.8 |
| 1 | 19 | 58 | 4 | 5 | 4 | 2 | 2 | 2 | 4 | 0 | 0 | 3.00×10$^{17}$ | 44.8 |
| 0 | 19 | 58 | 4 | 5 | 4 | 2 | 2 | 2 | 3 | 0 | 0 | substrate | |

| | | |
|---|---|---|
| layer-4 | (Lorentz-1) + SiO$_2$ | $d_4$ |
| layer-3 | (Lorentz-2) + silver | $d_3$ |
| layer-2 | (Lorentz-2) + ZrO$_2$ | $d_2$ |
| layer-1 | (Lorentz-2) | $d_1$ |
| | (Lorentz-1)   substrate | |

Figure 5. Multilayer optical model for the evaluation of spectroellipsometric data.

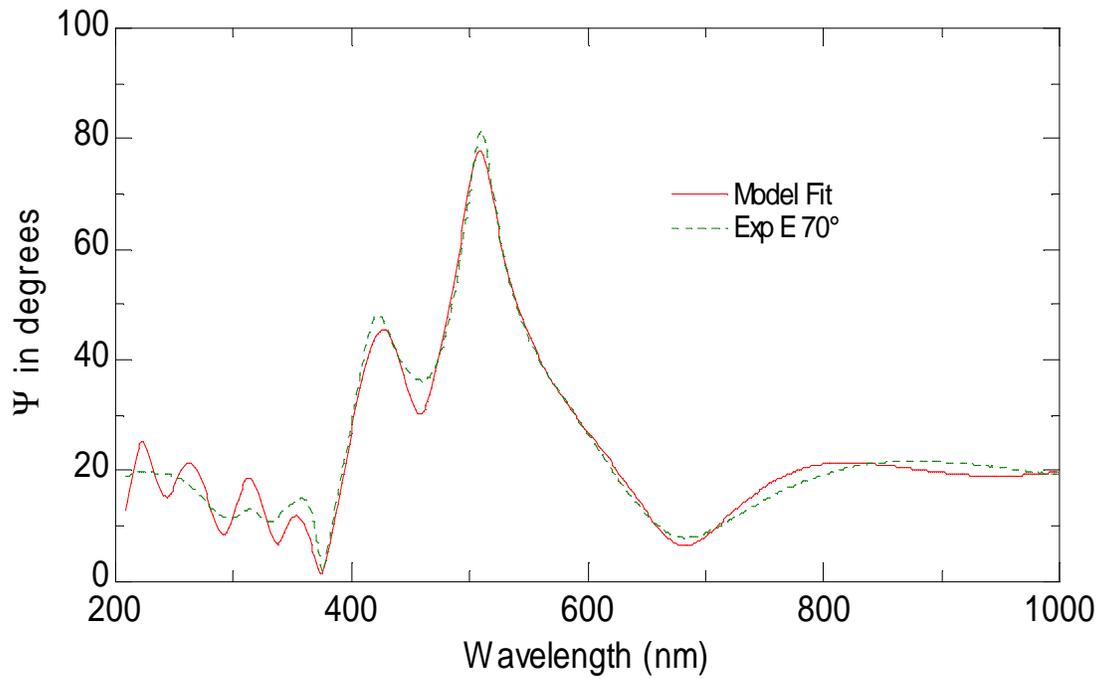

Figure 6. Measured and generated ellipsometric Ψ and Δ spectra for glazed ceramic pendant with greenish-gold metallic lustre decoration. A multilayer optical model with Lorentz oscillators and reference data from literature was applied. MSE=97.2.

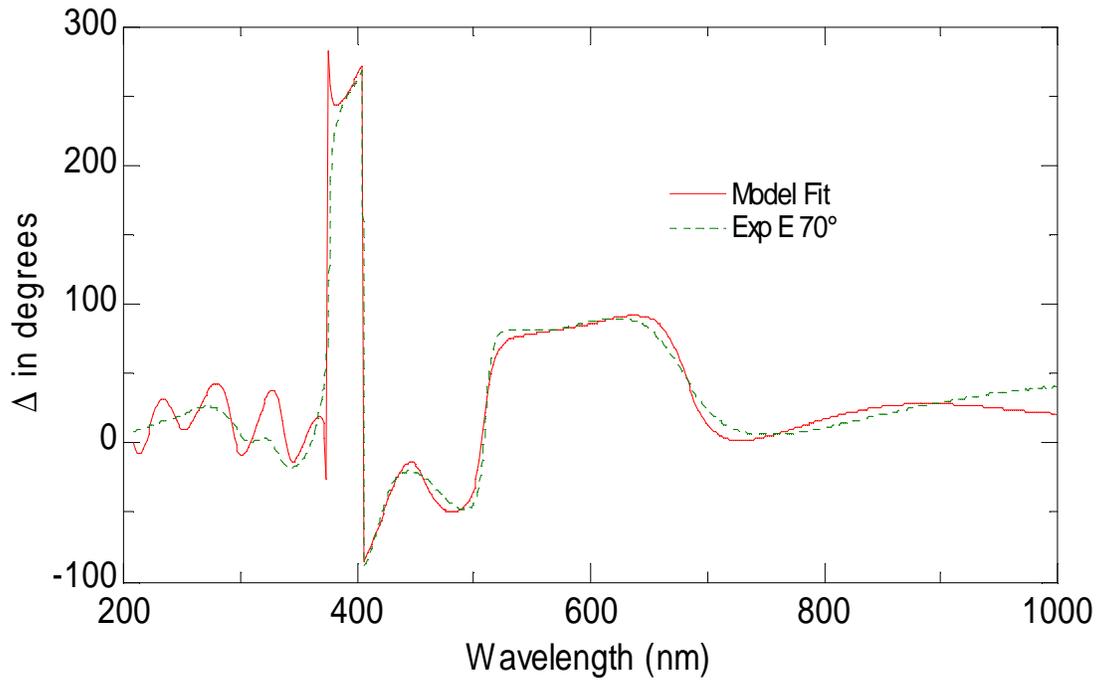

Figure 7. Measured and generated ellipsometric Δ spectra for glazed ceramic pendant with greenish-gold metallic lustre decoration. A multilayer optical model with Lorentz oscillators and reference data from literature was applied. MSE=97.2.

| | | |
|---|---|---|
| layer-4 | (Lorentz-1) + 82.7 ± 0.5 % SiO$_2$ | 88.4 ± 0.9 nm |
| layer-3 | (Lorentz-2) + 15.1 ± 1 % Ag | 9.8 ± 0.2 nm |
| layer-2 | (Lorentz-2) + 99.1 ± 0.2 % ZrO$_2$ | 85.1 ± 0.8 nm |
| layer-1 | (Lorentz-2) | 100 ± 3 nm |
| | (Lorentz-1) substrate | |

Figure 8. Multilayer optical model with the results of evaluation of spectroellipsometric data: the volume fractions and the layer thicknesses together with their confidence intervals are shown.